\documentclass[aps,pra,twocolumn,groupedaddress,superscriptaddress,showpacs]{revtex4}
\usepackage{graphicx}
\usepackage{amsmath,amssymb,amsfonts}
\usepackage{natbib}

\newcommand{\be}{\begin{equation}}
\newcommand{\ba}{\begin{eqnarray}}
\newcommand{\ee}{\end{equation}}
\newcommand{\ea}{\end{eqnarray}}

\begin{document}

\title{Local dynamics in high-order harmonic generation\\
 using Bohmian trajectories}

\author{J. Wu}
\affiliation{Department of Physics and Astronomy, University College
London, Gower Street, London WC1E 6BT, United Kingdom}

\author{B. B. Augstein}
\affiliation{Department of Physics and Astronomy, University College
London, Gower Street, London WC1E 6BT, United Kingdom}
\affiliation{School of Chemistry, University of Leeds, Leeds LS2 9JT, United Kingdom}

\author{C. Figueira de Morisson Faria}
\affiliation{Department of Physics and Astronomy, University College
London, Gower Street, London WC1E 6BT, United Kingdom}

\date{\today}

\begin{abstract}
We investigate high-order harmonic generation from a Bohmian-mechanical perspective, and find that
 the innermost part of the core, represented by a
single Bohmian trajectory, leads to the main contributions to the
high-harmonic spectra. Using time-frequency analysis, we associate this central Bohmian trajectory to an ensemble of unbound classical trajectories leaving and returning to the core, in agreement with the three step model. In the Bohmian scenario, this physical picture builds up non-locally near the core via the quantum mechanical phase of the wavefunction. This implies that the flow of the wavefunction far from the core alters the central Bohmian trajectory. We also show how this phase degrades in time for the peripheral Bohmian trajectories as they leave the core region.
\end{abstract}

\pacs{32.80.Rm, 42.50.Hz, 42.65.Ky}

\maketitle

\section{Introduction}
Since the past two decades, the concept of trajectories has been widely employed in strong-field physics.
 This concept has been introduced as early as 1993, together with the physical picture known as the ``three-step model" (TSM)
 \cite{Kulander:PRL:1993,corkum:PRL:1993}. According to this picture, strong-field phenomena are the result of the laser-induced recollision or recombination of an electron with its parent ion. The electron in question is freed by tunneling or multiphoton ionization, propagates in the continuum and is driven by the field towards the core. If recombination occurs, high order harmonic generation (HHG) takes place, while rescattering leads to above-threshold ionization (ATI), or nonsequential double and multiple ionization (NSDI,NSMI). This physical picture has become the cornerstone for describing these phenomena, and has demonstrated a high predictive power. This ranges from the plateau and the cutoff in HHG and ATI, to allowing a wide range of applications of HHG such as subfemtosecond pulses
\cite{Hentschel_2001,Keinberger_2004,Goulielmakis_2008}, and the
attosecond imaging of matter
\cite{Itatani_2004,Smirnova_2009,Vozzi_2011} (for reviews see
\cite{brabec:RMP:2000,ganeev:JPhysB:2007,midorikawa:JJAP:2011,altucci:JModOpt:2012}).

Initially, the TSM has been proposed in a classical framework. In these early models, an ensemble of electrons released in the continuum at different times within a field cycle was used in order to mimic the behavior of the electronic wavepacket. Depending on this time, the kinetic energy of each electron upon return would be different.  The maximal energy upon return gave the cutoff energy observed in experiments and in ab-initio computations, in which the time-dependent Schr\"odinger equation (TDSE) has been solved numerically \cite{tdse1}.
Soon thereafter, the above-mentioned physical picture has been
extracted from the expectation value of the dipole operator computed
in the strong-field approximation (SFA) \cite{lewenstein:PRA:1994}.
In the SFA, the time-dependent wavefunction is approximated by the
ground state and the continuum, which is taken as a superposition of
field-dressed plane waves. The electron trajectories associated with
the three-step model are then extracted from the phase of the
wavefunction using the steepest descent method. This has led to the
concept of ``quantum orbits", which is widespread in the
strong-field community \cite{Quantumorbit}. Furthermore, in recent
years other orbit-based approaches have been employed in the
strong-field context, such as the Volkov-eikonal approximation
\cite{Volkov-eikonal,Smirnova}, the Coulomb corrected strong-field
approximation
\cite{Bauer:JMO:2008,Bauer:PRL:2010,Bauer:PRA:2012}, the
adiabatic approximation \cite{Bondar:PRA:2009,Adiabatic:PRA:2012,Adiabatic_2},
the Herman-Kluk propagator \cite{Herman:CP:1984} and the coupled
coherent states method \cite{Dmitry:JCP:2000}.

It is also well known that ``cleaner" HHG spectra, with a large plateau and a well-defined cutoff, are obtained from the
expectation value of the dipole acceleration ${\bf a}(t) = \langle \Psi(t) | - \nabla V | \Psi(t) \rangle$, rather than from the dipole length in TDSE simulations \cite{burnett:PRA:1992,Krause:PRA:1992}. The dipole acceleration probes regions near the core, while the dipole length emphasizes regions closer to the integration boundaries \cite{Krause:PRA:1992}. This suggests that regions near the core, where the overlap between continuum and bound dynamics is likely to occur, are important to HHG. A legitimate question is, however, whether one can single out a specific region in the core as being the most relevant to HHG. Apart from that, one may ask how the above-stated overlap relates to the physical picture propagated by the TSM.

In this article, we investigate HHG using Bohmian mechanics \cite{bohm:PR:1952-1,holland-bk}. Bohmian trajectories are directly extracted from the TDSE, and act as ``tracer particles", i.e., they map the probability density flow in configuration space associated with the time-dependent wavefunction.
For that reason, both the time-dependent laser field and the binding potential are fully incorporated. Recently, Bohmian mechanics has been applied
to strong-field physics at a descriptive and interpretational level \cite{lai:EPJD:2009,botheron:PRA-2:2010,takemoto:JCP:2011,mompart} as well as a source for numerical algorithms
\cite{mompart,botheron:PRA-1:2010,Song_2012}. These papers essentially follow
the traditional scheme of considering a set of Bohmian trajectories
and comparing their statistics with the corresponding quantum
results. Here, in contrast, we will employ individual Bohmian trajectories in order to probe different regions in configuration space. Using a simplified, one-dimensional model, we show that (i) the Bohmian trajectory located in the innermost region of the core, in the vicinity of $x=0$, leads to high-order harmonic spectra with a plateau and a cutoff; (ii) this innermost trajectory may be associated with an ensemble of classical trajectories returning to its parent ion, according to the predictions of the TSM; (iii) in the Bohmian scenario, the picture related to the TSM builds up non-locally via the phase of the wavefunction. Any alterations in the flow of the wavefunction far from the core region will influence the central trajectory according to what is expected from the TSM. This will be exemplified by employing long- and short range potentials for which the core region is essentially the same, but for which the wavefunction propagation outside the core changes considerably.

This work is organized as follows. In Sec.~\ref{theory}, we provide the necessary theoretical background in order to understand the subsequent results. This includes a brief discussion on Bohmian trajectories (Sec.~\ref{Bohmian}) and on classical-ensemble models (Sec.~\ref{classical}). We also provide the windowed Fourier transforms employed to probe the phase of the wavefunction and of the central trajectory (\ref{windowed}). In Sec.~\ref{results}, we bring the outcome of our computations. We commence by discussing the Fourier spectra from the central trajectory, in comparison to the TDSE (\ref{Fourier}), and, subsequently, we analyze how this trajectory relates to those predicted in the three-step model (\ref{phase}). Finally, in Sec.~\ref{conclusions}, we will state our conclusions.
\section{Background}
\label{theory}
\subsection{Model}
In this work, we solve the TDSE in one spatial dimension. For linearly polarized fields, this suffices for a qualitative description of the system dynamics. We employ the length gauge and atomic units throughout.
The time-dependent Hamiltonian is given by %
\be
 H = - \frac{1}{2}\ \nabla^2 + V(x) - x E(t),
 \label{hamg}
\ee
where $E(t)$ denotes the driving field and $V(x)$ the binding potential (note that the minus sign in the last term arises from the electron
charge: in atomic units, $e=-1$).

The atomic potential $V(x)$ reads as
\be
 V(x) = - \frac{1}{\sqrt{x^2 + 1}}f(x) ,
 \label{potsc}
\ee
where  $f(x)$ is a function that will
determine the range of $V(x)$. If $f(x)=1$ throughout, $V(x)$ is a
long-range potential. If, however,
\be
f(x) = \left\{ \begin{array}{lcc}
  1 , & \quad & |x| < a_0 \\
  \displaystyle
  \cos^7 \left( \frac{\pi}{2} \frac{|x| - a_0}{L - a_0} \right) ,
   & \quad & a_0 \leq |x| \leq L \\
  0 , & \quad & |x| > L
  \end{array} \right. ,
 \label{tsc-mask}
\ee
the tail of $V(x)$ is truncated. Here, the parameters $a_0$ and $L$ have been chosen such that the core region is left practically unaltered, but the long tail of the potential is removed.
We shall refer to the long- and short range potentials as $V_{sc}(x)$ and $V_{tr}(x)$, respectively. For the parameters employed in this work, $V(x)$ supports several bound states. The ground-state energy is given by $\epsilon_0=-0.66995$ a.u.

The field is chosen to be a flat-top pulse $E(t) = E_0 g(t) \sin
(\omega t)$ of frequency $\omega$, with
\be
 g(t) = \left\{ \begin{array}{lcc}
  (t/\tau_0)  , & \quad
   0 \le t < \tau_{\rm on} \\
  1 , & \quad
   \tau_{\rm on} \le t < \tau_{\rm off} \\
 1 - (t - \tau_{\rm off})/\tau_{\rm on}
   , & \quad
   \tau_{\rm off} \le t \le \tau_f
 \end{array} \right. ,
 \label{pulse}
\ee
turned on and off in 2.25 cycles, i.e.,
$\tau_{\mathrm{on}}=2.25\tau_0$ and $\tau_{\mathrm{off}}=(2.25
+N)\tau_0$, where $\tau_0=2\pi/\omega$ is the field cycle, and
$\tau_f=\tau_{\mathrm{off}}+2.25 \tau_0$. Between turn on and turn
off, we consider $N=10$.

The time-dependent wavefunction $\Psi(x,t)$ is then obtained by solving the time-dependent Schr\"odinger equation
\be
 i \frac{\partial \Psi(x,t)}{\partial t} = H \Psi(x,t) ,
 \label{schro}
\ee
using the the fast Fourier transform (FFT) technique. The system is taken to be initially in its ground state, i.e., $\Psi(x,0)=\phi_0(x)$. For details on our method see Ref.~\cite{Wu}.

The expectation value of the dipole acceleration operator is computed as
\begin{equation}
a(t)=-\langle \Psi |dV(x)/dx|\Psi\rangle.
 \label{accel1}
\end{equation}
\subsection{Bohmian trajectories}
\label{Bohmian}
In order to construct the Bohmian trajectories, first the time-dependent wavefunction is written as $\Psi(x,t) = \rho^{1/2}(x,t)\ \! e^{iS(x,t)}$, with the
probability density $\rho$ and the phase $S$ being real-valued
functions of space and time.
This lead to the coupled differential equations,
\be
 \frac{\partial \rho}{\partial t} + \nabla \cdot {\bf J} = 0 ,
 \label{cont}
\ee
where ${\bf J} = \rho \nabla S$ is the usual quantum probability
current density,
and
\be
 \frac{\partial S}{\partial t} + \frac{(\nabla S)^2}{2} + V + Q = 0 ,
 \label{qHJ}
\ee
where
\be
 Q(x,t) = - \frac{1}{2} \frac{\nabla^2 \rho^{1/2}}{\rho^{1/2}}
  = - \frac{1}{4} \left[ \frac{\nabla^2 \rho}{\rho}
   - \frac{1}{2} \left(\frac{\nabla \rho}{\rho}\right)^2 \right].
\ee
 Eq.~(\ref{cont}) is known as the continuity equation, and Eq.~(\ref{qHJ}) is known as the quantum Hamilton-Jacobi equation.  In Eq.~(\ref{qHJ}), $Q(x,t)$ is the quantum potential and $S$ is the equivalent of the
classical action.

Bohmian trajectories are
obtained \cite{note-wyatt} after integrating the (also real-valued)
guidance equation
\be
 \dot{x} = \nabla S = \frac{\bf J}{\rho}
   = \frac{1}{2i}
  \left(\frac{\Psi^* \nabla \Psi - \Psi \nabla \Psi^*}{|\Psi|^2}\right),
 \label{eom}
\ee
 at each time-step, i.e., once
$\Psi(x,t)$ is known.

\subsection{Classical-ensemble computations}
\label{classical}
 In order to compare Bohmian and classical trajectories, we solve the classical equations of motion of an ensemble of electrons, which are released in the laser field at a time $t_0$. For each electron,
\be
\ddot{x}=F(x,t)
\label{Newton}
\ee
 both in the presence and in the absence of the soft-core potential. In the former and the latter case, $F(x,t)=E(t)-dV(x)/dx$ and $F(x,t)=E(t)$, respectively.  The initial release time $t_0$ is then varied within a monochromatic field given by $E(t)$ with $g(t)=1$ in Eq.~(\ref{pulse}).  Only a subset of the trajectories obtained will return to the core, depending on the time at which the electrons  are released into the field. For a monochromatic field this will occur only for times $t_0>0.25\tau_0+n\tau_0/2$, i.e., after the peak-field times, and up to $t_0\simeq 0.4\tau_0+n\tau_0/2$, i.e., somewhat before the crossing.

If the potential is absent, integrating Eq.~(\ref{Newton}) once and twice gives
\begin{equation}
\dot{x}=\frac{E_0}{\omega}\Big(\cos(\omega t_0)-\cos(\omega
t)\Big)+v_0,\label{classical1}
\end{equation}
and
\begin{eqnarray}
x&=&\frac{E_0}{\omega^2}\Big((t-t_0)\cos(\omega t_0)-\sin(\omega
t)
+\sin(\omega t_0)\Big)\\ \notag&&+v_0(t-t_0)+x_0, \label{PosEq}
\end{eqnarray}
respectively. In the above-stated equations, the initial position and velocity of the electron are $x_0$ and $v_0$, respectively. In this case, we choose $v_0=0$ and $x_0=0$.

 If the soft-core potential is included, Eq.~(\ref{Newton}) is rewritten as the coupled first order differential equations,
\begin{equation}
v=\dot{x}
\end{equation}
and
\begin{equation}
\dot{v}=E_0\sin(\omega t)-\nabla V(x),
\end{equation}
which are solved employing the fourth order Runge Kutta method. Note, however, that solving these equations with the same initial conditions as in the potential-free case, that is with $x_0=0$ and $v_0=0$ leads to a series of bound trajectories, whose kinetic energy is very low and which are not appropriate for the comparison one is willing to perform. Hence, we have placed the electrons initially at $x_0=0$ but with velocity $v_0$ such that $v^2_0/2=-V(x_0)$. This gives $v_0=\pm \sqrt{2}$ for the potentials employed in this work. These trajectories will exhibit unbound dynamics. Similar dynamics may be obtained by assuming that the electrons in the ensemble leave with vanishing velocity $v_0=0$, but are initially located at a turning point, i.e., at $x_0$ such that $F(x_0,t_0)=0$.

\subsection{Windowed Fourier transforms}
\label{windowed} Here, we will calculate the standard Fourier spectrum of
the Bohmian trajectories $x_{B}(t)$ and of the dipole acceleration
$a(t)$, which is given by
\be
 I(\Omega) = \left\arrowvert \int h(t)\ \! e^{i\Omega t}
   dt \right\arrowvert^2,
 \label{fourier0}
\ee
where $h(t)$ generically denotes either $x_{B}(t)$ or $a(t)$ and the integral is the standard Fourier transform $a_F(\Omega)$. Apart from that, we
will also employ windowed Fourier transforms, in the form of
\begin{equation}
a_G(\Omega, t^{\prime})=\int h(t)\exp[-(t-t^{\prime})^2/(2\sigma^2)]\exp(i\Omega
t)dt, \label{Gabor}
\end{equation}
to introduce temporal resolution in the HHG spectra. Eq.~(\ref{Gabor}) is known as the Gabor transform, and has been widely used to extract temporal information from the TDSE (see, e.g.,
Refs.~\cite{timefrequency2,FDS1997,Belgium1998,timefrequency1}; or for recent references Ref.~\cite{Ruggenthaler_2008,Lein:PRA:2010,Ciappina_1_2012,Ciappina_2_2012}). If $\sigma \rightarrow \infty$ the standard Fourier transform is recovered and all temporal information is lost.
\section{Results}
\label{results}

\subsection{High-harmonic spectra}
\label{Fourier}

\begin{figure}[t]
\noindent\hspace*{-0.5cm}\includegraphics[width=10cm]{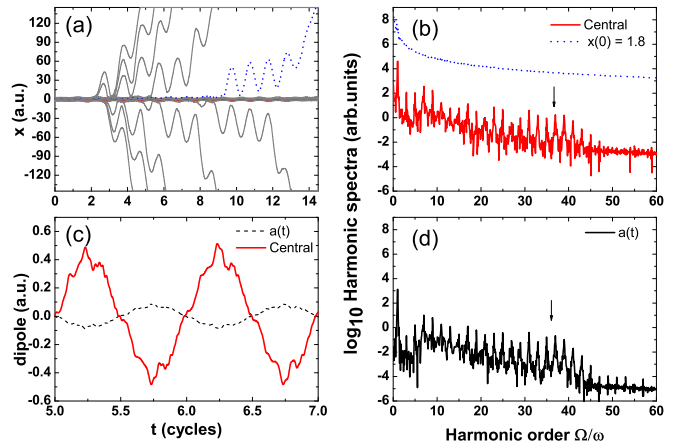}
\vspace{-0.8cm}
 \caption{\label{fig1}(Color online)
  Panel (a): Set of Bohmian trajectories obtained using a long-range softcore potential $V_{sc}(x)$ in a flat-top pulse with peak field strength $E_0 =0.075$ a.u. and frequency $\omega = 0.057 $ a.u. (Keldysh parameter $\gamma=0.88$), with initial ranging from $x(0)=-3$ a.u. to $x(0)=3$ a.u. Panel (b): High order harmonic spectra from the central Bohmian trajectory $x_c(t)$ ($x(0) = 0$)(red solid line) and from a peripheral Bohmian trajectory starting at $x(0) = 1.8$ a.u. (blue dotted line). Panel (c): a blow-up of the central trajectory over two cycles of the laser field (red solid line), together with the expectation value of the dipole acceleration computed from the TDSE (black dashed line); note that, formally, the central Bohmian trajectory $x_c(t)$ is equivalent to the time-dependent dipole length computed using only the innermost part of the TDSE wavefunction. Panel (d): Power spectra from the dipole acceleration computed from the TDSE, plotted using the same scale as in panel (b) to facilitate a direct comparison. The cutoff frequency according to the three-step model is indicated by the arrows in panels (b) and (d). }
  \end{figure}

We first study the different subsets of Bohmian trajectories and their power spectra, displayed in Fig.~\ref{fig1}.  These trajectories illustrate the flow of the probability density in configuration space. Throughout, the driving-field parameters are chosen such that the system is in the tunneling regime. Unless otherwise stated, we will consider the long-range softcore potential $V_{sc}(x)$.

In Fig.~\ref{fig1}(a), one may identify two distinct subsets of Bohmian trajectories: those that oscillate within the core region, and those that oscillate far from the core until they eventually leave. In Fig.~\ref{fig1}(b), we display the spectra obtained from the central Bohmian trajectory, i.e., that starting at $x(0)=0$, and from a trajectory starting at a few atomic units
from the core ($x(0)=1.8$ a.u.). The spectrum of the peripheral trajectory consists of a smooth, monotonically decaying background and a small signature around
the fundamental, $\Omega = \omega$, with no harmonic peaks. In contrast, the central Bohmian trajectory ($x(0) = 0$) gives us a
clear high-order harmonic spectrum with a large plateau followed by
a sharp cutoff located at $|\epsilon_0|+3.17U_p$. As the initial condition $x(0)$ of a specific Bohmian trajectory gets further away
from $x=0$, not only does the power spectrum of the corresponding
trajectory lose the plateau and the cutoff, but it also
gains intensity. Hence, if an average of Bohmian trajectories across the whole configuration space is taken in order to compute the spectra, both the plateau and the cutoff will be obscured. This problem is also encountered when computing HHG spectra using the length form of the dipole operator and it is overcome either by using the dipole acceleration or numerical filters in frequency space. The former emphasizes the core region, and the latter change the flow of the wavefunction in real time by forcing the probability density to return to the core. For comparison, the power spectrum from the dipole acceleration obtained from the TDSE is displayed in Fig.~\ref{fig1}(d).

In Fig.~\ref{fig1}(c), we have a closer look at the dipole acceleration and the central Bohmian trajectory. The figure shows that both not only follow the field, but exhibit a series of high-frequency oscillations. These oscillations are not present in peripheral Bohmian trajectories.  A noteworthy feature is that, on average, the distance in time between
adjacent peaks is around $0.03$ times the length of a cycle. This corresponds to a typical frequency of about $35\omega$, which is roughly the cutoff frequency. Similar oscillations have also been identified in the dipole acceleration, both in TDSE computations \cite{Knight:PRA:1996}, or employing other orbit-based methods \cite{Adiabatic:PRA:2012,Rost:PRL:1999,Carlos:PRA:2012,Jie:Conference:2012}. In fact, early studies have identified these oscillations as paramount for obtaining a plateau and a cutoff, together with the strong localization of the dipole acceleration in configuration space \cite{Knight:PRA:1996}. They have been associated with the interference between the outgoing and incoming parts of the electronic wavepacket, which overlap near the core. Recently, similar arguments have been put across using the adiabatic approximation \cite{Adiabatic:PRA:2012,Adiabatic_2}. Therein, it has been shown that part of the electronic wavefunction exhibits a highly oscillating phase. This phase may be associated with the classical action of an electron leaving and returning to the core, and contributes to the action $S(x,t)$ defining the Bohmian trajectories. Furthermore, studies employing the Herman Kluk propagator \cite{Rost:PRL:1999,Carlos:PRA:2012} and the
coupled-coherent states method \cite{Jie:Conference:2012} have found that this highly oscillating structure is related to the quantum interference between different types of electron trajectories returning to the core.
\subsection{Time-frequency analysis}
\label{phase}
Next, we wish to address the question of how the Bohmian trajectories compare to the classical trajectories of an electron in a strong laser field. Furthermore, we would like to have a closer look at the phase of the time-dependent wavefunction. Specifically, we will assess whether information may be transferred nonlocally via this phase to the central trajectory by altering the flow of the wavefunction far from the core.

With that purpose in mind, we truncate the long-range potential
according to Eq.~(\ref{tsc-mask}) so that the core region is kept practically
unaltered, i.e., its field-free eigenerergies are very close to
those of the long-range potential, but the long-range tail of the soft-core potential is eliminated. Below, in Table \ref{tab1}, we give the bound-state energies for the two potentials.
 \begin{table}
 \caption{\label{tab1}
  Eigenvalues for the long-range soft-core and the
  truncated soft-core potential, for which $a_0=5.0$ and $L=50$. Note that, in principle, the number of
  eigenstates supported by the long-range potential is infinity; in our calculations, though, we obtain a finite number of them because of the boundaries of the grid we are using to solve
  the TDSE. All quantities are given in a.u.}
 \begin{tabular}{c p{.3cm} c p{.3cm} c p{.3cm} c}
  \hline\hline
   \ $n$ & & untruncated & & truncated  \\ \hline
   \  0  & & -0.66995    & & -0.66995  \\
   \  1  & & -0.27508    & & -0.27503  \\
   \  2  & & -0.15158    & & -0.15059  \\
   \  3  & & -0.09276    & & -0.08714 \\
   \  4  & & -0.06358    & & -0.05013  \\
   \  5  & & -0.04552    & & -0.02390  \\
   \  6  & & -0.03462    & & -0.00754   \\
   \  $\vdots$ & & $\vdots$    \\
   \  14  & & -0.00826     \\
   \  15  & & -0.00707      \\
   \  16  & & -0.00670     \\
  \hline\hline
 \end{tabular}
\end{table}

\begin{figure}[t]
 \begin{center}
 \includegraphics[width=8.5cm]{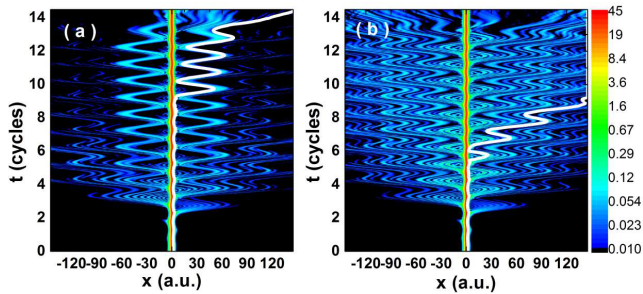}
 \caption{\label{fig2}(Color online)
  Time-evolution of the probability density distribution $|\Psi(x,t)|^2$ in configuration space for the long-range
  potential $V_{sc}(x)$[panel (a)] and the short-range potential $V_{tr}(x)$ [panel(b)] and the same laser parameters used in Fig.~\ref{fig1}. For clarity, the peripheral Bohmian trajectories starting at $x(0)=1.8$ a.u. employed in the time frequency maps of Fig.~\ref{fig3} are highlighted as the white curves in the figure. The maps in panels (a) and (b) have been multiplied by 100.}
 \end{center}
\end{figure}

For the sake of clarity, in Fig.~\ref{fig2} we display the probability density flow for both potentials. The figure shows that the outward flow is larger for the short-range potential $V_{tr}(x)$[Fig.~\ref{fig2}(b)], compared to its long range counterpart $V_{sc}(x)$ [Fig.~\ref{fig2}(a)]. This is due to the fact that the Coulomb tail restricts this flow. This confinement is absent in the short-range case.

\begin{figure}[t]
\noindent\hspace*{-0.8cm}
\includegraphics[width=9.5cm]{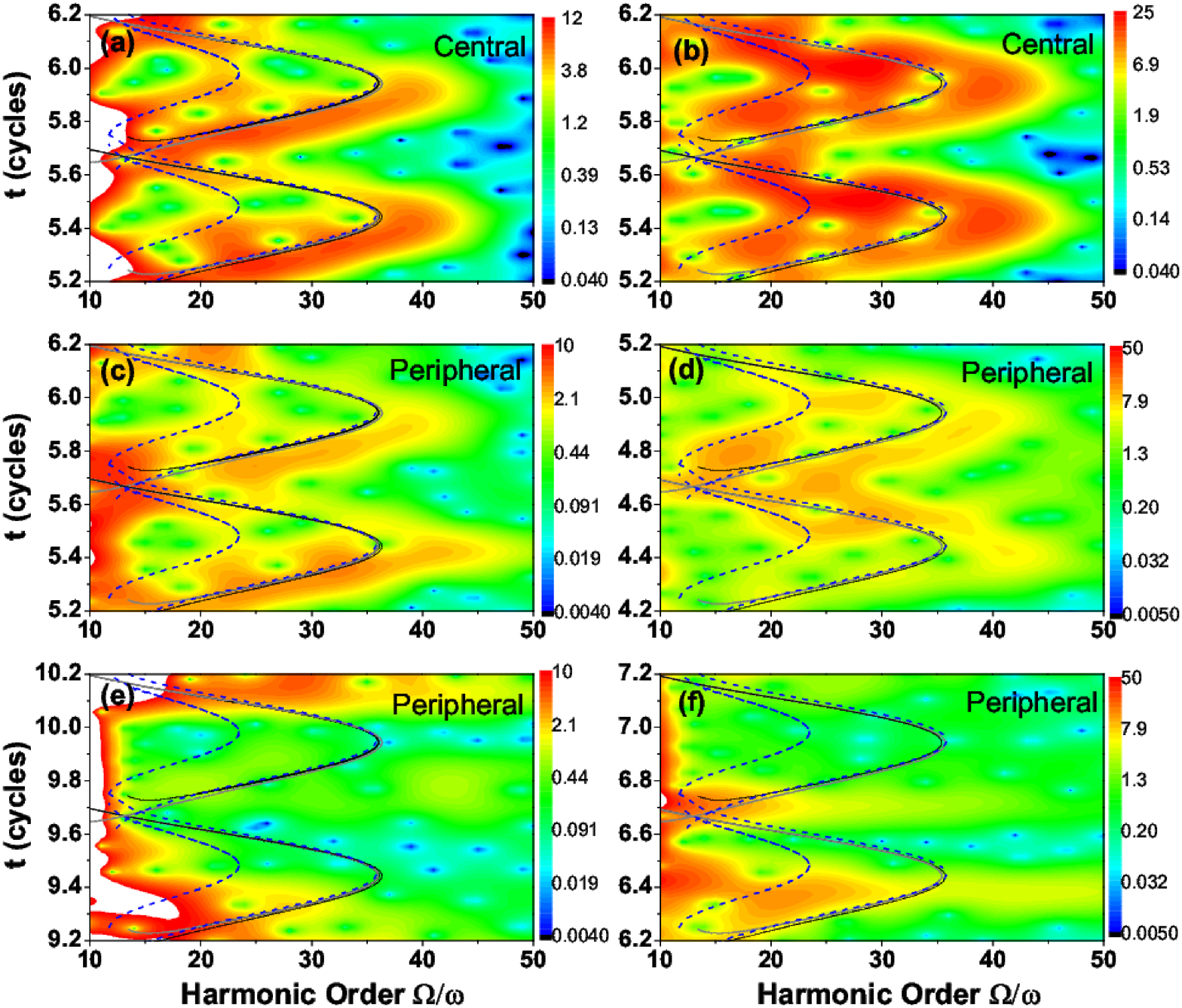} \vspace*{-0.2cm}
 \caption{\label{fig3}(Color online)
  Time frequency maps as functions of the harmonic order computed using the Gabor transform (\ref{Gabor}), for the central Bohmian trajectory [panels (a) and (b)], and the peripheral trajectory starting at $x(0)=1.8$a.u. [panels (c) to (f)] for a one-dimensional atomic model in a trapezoidal field of frequency
  $ \omega = 0.057$ a.u. and intensity $E_0 = 0.075$ a.u..  In panels (a), (c) and (e) the long range potential $V_{sc}(x)$ has been used, while in panels (b), (d) and (f) the truncated potential $V_{tr}(x)$ has been considered. The solid and the blue dashed arches in the figure are related to the outcome of the classical-trajectory computations with and without binding potential, respectively. In the classical-trajectory computations with binding potential, we considered that the electron ensemble was released with escape velocity $v_0$ according to Sec.~\ref{classical}. The black thin lines account for the solutions obtained using positive values of $v_0$, while the gray thick lines correspond to the solutions obtained for negative $v_0$.
   In the middle panels we consider a temporal window for which the peripheral trajectory in question is still close to the core (between the 5th and the 6th cycle for $V_{sc}(x)$ and between the 4th and the 5th cycle for $V_{tr}(x)$), while in the lower panels we take a time interval for which it has left this region (between the 9th and the 10th cycle for $V_{sc}(x)$ and between the 6th and the 7th cycle for $V_{tr}(x)$). The field parameters are the same as in the previous figures. The maps in panels (a) and (b) have been multiplied by 100, while those in the remaining panels have been multiplied by 10 to facilitate a better comparison.}
\end{figure}

In order to extract such trajectories from the phase of the wavefunction, we will construct time-frequency maps employing the windowed Fourier transform (\ref{Gabor}). Throughout, we use the same window function as in Ref.~\cite{Lein:PRA:2010}, i.e., $\sigma=1/(3\omega)$. In Fig.~\ref{fig3}, we show these time frequency maps for the central and the peripheral Bohmian trajectories highlighted in Fig.~\ref{fig2}. The left and the right panels are related to the long- and short- range potential $V_{sc}$ and $V_{tr}$, respectively.

Overall, the time-frequency maps associated with the central Bohmian trajectory, shown in Figs.~\ref{fig3}(a) and (b), are in full agreement with the three-step model, regardless of whether the long- or the short-range potential was taken. Indeed, these maps exhibit a series of arches, which correspond to the classical return times of an electron ensemble leaving the core, propagating in the continuum and recombining with their parent ion. For clarity, these return times are indicated by the curves in the figure. Each point in these curves gives the return time of a classical electron in the field for a specific harmonic energy, and thus determine a classical orbit together with the start time $t_0$ \cite{Antoine_1996}.  Some discrepancies, however, occur depending on whether the binding potential has been neglected or included in the classical computations. In the latter case, there are more solutions for the return condition $x(t)=0$, which depend on whether the initial electron velocity is on the same direction or opposite to the field. These solutions have been recently discussed in Ref.~\cite{Hostetter_2010}, in a Coulomb-corrected SFA model. The lower parts of the arches correspond to the so-called ``short" trajectories, along which each classical electron returns before the field crossing, and the upper parts of the arches correspond to the ``long" trajectories, for which it returns after the crossing. In the time-frequency profiles computed for the long-range potential, the lower parts of the arches are more intense. This indicates that the contributions of the short classical trajectories are dominant. Apart from that, one also observes faint second arches, extending up to harmonic energy of approximately $|\epsilon_0|+1.5U_p$. These arches are related to even longer classical electron trajectories, with excursion times $t-t_0$ of the order of one and a half cycle.

Fig.~\ref{fig3} also shows that, if the flow of the wavefunction far from the core is altered by truncating the tail of the long-range potential, the time-frequency maps obtained for the central Bohmian trajectory will be influenced. Indeed, the upper part of the arches in these maps will become more intense. This implies that contributions from the longer classical orbits will become more prominent, in agreement with what has been observed in the literature \cite{Lein:PRA:2010}. Hence, the phase
information contained in the innermost part of the wave function will be changed non-locally by altering the flow far away from the core.

The time-frequency profiles of the peripheral Bohmian trajectory starting at $x(0)=1.8$ a.u., depicted in the remaining panels of Fig.~\ref{fig3}, behave in a rather different way. We have chosen the initial and final times so that the Bohmian trajectories in question are still within, or have just left the core region. An illustration of how the probability flow behaves at such times is provided in Fig.~\ref{fig2}, for comparison. In case the trajectory is still close the core, arch-like structures may be identified in the time-frequency maps, as shown in Figs.~\ref{fig3}(c) and (d), which, once more, corresponds to the return times predicted by the TSM. Nonetheless, these structures are more blurred than those observed for the central trajectory. As the Bohmian trajectories move away from the core region, these structures degrade very quickly, and the agreement with the TSM is lost. This can be observed in Figs.~\ref{fig3}(d) and (f).
\section{Conclusions}
\label{conclusions}
Summarizing, Bohmian trajectories show that the main contribution to
the HHG spectrum arises from the most internal part of the wave
function. Indeed, a single Bohmian trajectory
contains all the information necessary to obtain the HHG spectrum,
namely the trajectory that starts at $x_0=0$. This is a stronger statement than that provided by the dipole acceleration: By
using the acceleration, one may conclude that the overlap between
the continuum and bound part of the wavefunction near the core
region are important. Here, we show that the part of the
wavefunction located in the immediate vicinity of $x=0$ provides the
HHG spectrum.
We have chosen several driving-field intensities,
frequencies, pulse shapes and binding potentials (not only those
presented in this work) in order to corroborate these results are general. Some of these results have been included elsewhere \cite{Wu}.

In order to understand the above-stated results, one should keep in mind that a Bohmian trajectory is a non-local entity, i.e., it functions much more like a ``slice" of the wavefunction than like a trajectory in the classical sense. Only for coherent states and very specific ranges of the Mandel parameter may a Bohmian trajectory be associated with a classical trajectory \cite{Andreas}. In general, however, this is not the case. In fact, a Bohmian trajectory evolves under the action of the wave function, which not only encompasses local information about the space variations of the potential function, but also about global changes of the quantum phase. This implies that a Bohmian trajectory may be localized in the innermost part of the core and still contain bound and continuum dynamics. Any change in the wavefunction, be it far or close to the core region, will be transmitted non-locally to the central trajectory via its phase.

This is consistent with the fact that, in quantum-mechanical and semiclassical models, the trajectories related to the TSM are always extracted from the phase of the wavefunction, i.e., from the action. This holds both in the SFA, when these trajectories are obtained using the steepest descent method \cite{lewenstein:PRA:1994}, or when other methods are used, such as the Herman Kluk propagator \cite{Rost:PRL:1999,Carlos:PRA:2012} or the adiabatic approximation \cite{Adiabatic:PRA:2012,Adiabatic_2}. Our time-frequency maps support the fact that this phase behaves as an ensemble of unbound classical trajectories following the predictions of the TSM. Any alterations in the flow of the wavefunction far from the core will affect how this phase builds up. Furthermore, our results confirm the well-known fact that, spatially, HHG takes place at the core. In fact, time-frequency analysis of peripheral Bohmian trajectories illustrate the degradation of the above-mentioned profile when the probability density flow distances itself from the core region.

Finally, our studies also illustrate why the SFA works so well. The SFA
reduces the influence of the core to a single point, i.e., $x=0$, and approximates the continuum by Volkov waves.
This is a good approximation, because the most relevant part of
$\Psi(x,t)$ for HHG is strongly localized. Further evidence for this similarity has been provided by us in Ref.~\cite{Wu}, in which we show that the time profile of the central Bohmian trajectory overestimates the influence of the long TSM trajectory, in comparison with the TDSE. This over-enhancement is also known to occur in the SFA \cite{Perez_2010,Gaarde:PRA:2002}.

\textbf{Acknowledgements:} This work was supported in part by the UK EPSRC and the CSC/BIS. We are greatly indebted to A. S. Sanz for his collaboration in the initial stages of this project. J. W. and C. F. M. F. would like to thank A. Fring, C. Zagoya and X. Lai for useful discussions.


\begin{thebibliography}{21}
\expandafter\ifx\csname natexlab\endcsname\relax\def\natexlab#1{#1}\fi
\expandafter\ifx\csname bibnamefont\endcsname\relax
  \def\bibnamefont#1{#1}\fi
\expandafter\ifx\csname bibfnamefont\endcsname\relax
  \def\bibfnamefont#1{#1}\fi
\expandafter\ifx\csname citenamefont\endcsname\relax
  \def\citenamefont#1{#1}\fi
\expandafter\ifx\csname url\endcsname\relax
  \def\url#1{\texttt{#1}}\fi
\expandafter\ifx\csname urlprefix\endcsname\relax\def\urlprefix{URL }\fi
\providecommand{\bibinfo}[2]{#2}
\providecommand{\eprint}[2][]{\url{#2}}


\bibitem[{\citenamefont{Corkum}(1993)}]{Kulander:PRL:1993}
\bibinfo{author}{\bibfnamefont{K.~C.} \bibnamefont{Kulander} \bibnamefont{{\it et al.}}},
  \bibinfo{journal}{Phys. Rev. Lett.} \textbf{\bibinfo{volume}{70}},
  \bibinfo{pages}{1599} (\bibinfo{year}{1993}).

\bibitem[{\citenamefont{Corkum}(1993)}]{corkum:PRL:1993}
\bibinfo{author}{\bibfnamefont{P.~B.} \bibnamefont{Corkum}},
  \bibinfo{journal}{Phys. Rev. Lett.} \textbf{\bibinfo{volume}{71}},
  \bibinfo{pages}{1994} (\bibinfo{year}{1993}).


\bibitem[{\citenamefont{Hentschel et~al.}(2001)}]{Hentschel_2001}
\bibinfo{author}{\bibfnamefont{M.}~\bibnamefont{Hentschel} \bibnamefont{{\it et al.}}},
  \bibinfo{journal}{Nature} \textbf{\bibinfo{volume}{414}},
  \bibinfo{pages}{509} (\bibinfo{year}{2001}).



\bibitem[{\citenamefont{Kienberger et~al.}(2004)}]{Keinberger_2004}
\bibinfo{author}{\bibfnamefont{R.}~\bibnamefont{Keinberger} \bibnamefont{{\it et al.}}},
  \bibinfo{journal}{Nature} \textbf{\bibinfo{volume}{427}},
  \bibinfo{pages}{817} (\bibinfo{year}{2004}).


\bibitem[{\citenamefont{Goulielmakis et~al.}(2008)}]{Goulielmakis_2008}
\bibinfo{author}{\bibfnamefont{E.}~\bibnamefont{Goulielmakis} \bibnamefont{{\it et al.}}},
  \bibinfo{journal}{Science} \textbf{\bibinfo{volume}{320}},
  \bibinfo{pages}{1614} (\bibinfo{year}{2008}).

\bibitem[{\citenamefont{Itatani et~al.}(2009)}]{Itatani_2004}
\bibinfo{author}{\bibfnamefont{J.}~\bibnamefont{Itatani} \bibnamefont{{\it et al.}}},
  \bibinfo{journal}{Nature} \textbf{\bibinfo{volume}{432}},
  \bibinfo{pages}{867} (\bibinfo{year}{2004}).

\bibitem[{\citenamefont{Smirnova et~al.}(2009)}]{Smirnova_2009}
\bibinfo{author}{\bibfnamefont{O.}~\bibnamefont{Smirnova} \bibnamefont{{\it et al.}}},
  \bibinfo{journal}{Nature} \textbf{\bibinfo{volume}{460}},
  \bibinfo{pages}{972} (\bibinfo{year}{2009}).

\bibitem[{\citenamefont{Vozzi et~al.}(2011)}]{Vozzi_2011}
\bibinfo{author}{\bibfnamefont{C.}~\bibnamefont{Vozzi} \bibnamefont{{\it et al.}}},
  \bibinfo{journal}{Nat. Phys.} \textbf{\bibinfo{volume}{7}},
  \bibinfo{pages}{822} (\bibinfo{year}{2011}).

\bibitem[{\citenamefont{Brabec and Krausz}(2000)}]{brabec:RMP:2000}
\bibinfo{author}{\bibfnamefont{T.}~\bibnamefont{Brabec}} \bibnamefont{and}
  \bibinfo{author}{\bibfnamefont{F.}~\bibnamefont{Krausz}},
  \bibinfo{journal}{Rev. Mod. Phys.} \textbf{\bibinfo{volume}{72}},
  \bibinfo{pages}{545} (\bibinfo{year}{2000}).

\bibitem[{\citenamefont{Ganeev}(2007)}]{ganeev:JPhysB:2007}
\bibinfo{author}{\bibfnamefont{R.~A.} \bibnamefont{Ganeev}},
  \bibinfo{journal}{J. Phys. B}
  \textbf{\bibinfo{volume}{40}}, \bibinfo{pages}{R213}
  (\bibinfo{year}{2007}).

\bibitem[{\citenamefont{Midorikawa}(2011)}]{midorikawa:JJAP:2011}
\bibinfo{author}{\bibfnamefont{K.}~\bibnamefont{Midorikawa}},
  \bibinfo{journal}{Jpn. J. Appl. Phys.} \textbf{\bibinfo{volume}{50}},
  \bibinfo{pages}{090001} (\bibinfo{year}{2011}).

\bibitem[{\citenamefont{Altucci et~al.}(2011)\citenamefont{Altucci, Tisch, and
  Velotta}}]{altucci:JModOpt:2012}
\bibinfo{author}{\bibfnamefont{C.}~\bibnamefont{Altucci}},
  \bibinfo{author}{\bibfnamefont{J.~W.~G.} \bibnamefont{Tisch}},
  \bibnamefont{and} \bibinfo{author}{\bibfnamefont{R.}~\bibnamefont{Velotta}},
  \bibinfo{journal}{J. Mod. Opt.} \textbf{\bibinfo{volume}{58}},
  \bibinfo{pages}{1585} (\bibinfo{year}{2011}).
\bibitem{tdse1} J. L. Krause, K. J. Schafer and K. C. Kulander, Phys. Rev. Lett. \textbf{68}, 3535 (1992).


\bibitem[{\citenamefont{Lewenstein et~al.}(1994)\citenamefont{Lewenstein,
  Balcou, Ivanov, L'Huillier, and Corkum}}]{lewenstein:PRA:1994}
\bibinfo{author}{\bibfnamefont{M.}~\bibnamefont{Lewenstein} \bibnamefont{{\it et al.}}},
 \bibinfo{journal}{Phys. Rev. A}
  \textbf{\bibinfo{volume}{49}}, \bibinfo{pages}{2117} (\bibinfo{year}{1994}).


\bibitem[{\citenamefont{Quantumorbit}(1993)}]{Quantumorbit}
\bibinfo{author}{\bibfnamefont{P.} \bibnamefont{Sali$\grave{e}$res} \bibnamefont{{\it et al.}}},
  \bibinfo{journal}{Science} \textbf{\bibinfo{volume}{292}},
  \bibinfo{pages}{902} (\bibinfo{year}{2001}).

\bibitem[{\citenamefont{Volkov-eikonal}(2008)}]{Volkov-eikonal}
\bibinfo{author}{\bibfnamefont{O.} \bibnamefont{Smirnova} \bibnamefont{{\it et al.}}},
  \bibinfo{journal}{Phys. Rev. A} \textbf{\bibinfo{volume}{77}},
  \bibinfo{pages}{033407} (\bibinfo{year}{2008}).

\bibitem[{\citenamefont{Smirnova}(2008)}]{Smirnova}
\bibinfo{author}{\bibfnamefont{O.} \bibnamefont{Smirnova} \bibnamefont{{\it et al.}}},
  \bibinfo{journal}{J. Phys. B} \textbf{\bibinfo{volume}{39}},
  \bibinfo{pages}{S307} (\bibinfo{year}{2006}).

\bibitem[{\citenamefont{Bauer}(2008)}]{Bauer:JMO:2008}
  \bibinfo{author}{\bibfnamefont{S.~V.}~\bibnamefont{Propuzhenko}} \bibnamefont{and}
  \bibinfo{author}{\bibfnamefont{D.}~\bibnamefont{Bauer}},
  \bibinfo{journal}{J. Mod. Opt.} \textbf{\bibinfo{volume}{55}},
  \bibinfo{pages}{2573} (\bibinfo{year}{2008}).

\bibitem[{\citenamefont{Bauer}(2010)}]{Bauer:PRL:2010}
\bibinfo{author}{\bibfnamefont{T. M.} \bibnamefont{Yan} \bibnamefont{{\it et al.}}},
  \bibinfo{journal}{Phys. Rev. Lett.} \textbf{\bibinfo{volume}{105}},
  \bibinfo{pages}{253002} (\bibinfo{year}{2010}).

\bibitem[{\citenamefont{Bauer}(2012)}]{Bauer:PRA:2012}
  \bibinfo{author}{\bibfnamefont{T.~M.}~\bibnamefont{Yan}} \bibnamefont{and}
  \bibinfo{author}{\bibfnamefont{D.}~\bibnamefont{Bauer}},
  \bibinfo{journal}{Phys. Rev. A} \textbf{\bibinfo{volume}{86}},
  \bibinfo{pages}{053403} (\bibinfo{year}{2012}).

\bibitem[{\citenamefont{Bondar}(2009)}]{Bondar:PRA:2009}
\bibinfo{author}{\bibfnamefont{D. I.} \bibnamefont{Bondar} \bibnamefont{{\it et al.}}},
  \bibinfo{journal}{Phys. Rev. A} \textbf{\bibinfo{volume}{79}},
  \bibinfo{pages}{065401} (\bibinfo{year}{2009}).

\bibitem[{\citenamefont{Adiabatic}(2012)}]{Adiabatic:PRA:2012}
  \bibinfo{author}{\bibfnamefont{O.~I.}~\bibnamefont{Tolstikhin}} \bibnamefont{and}
  \bibinfo{author}{\bibfnamefont{T.}~\bibnamefont{Morishita}},
  \bibinfo{journal}{Phys. Rev. A} \textbf{\bibinfo{volume}{86}},
  \bibinfo{pages}{043417} (\bibinfo{year}{2012}).
\bibitem{Adiabatic_2} Y. Okajima, O. I. Tolstikhin, and T. Morishita, Phys. Rev. A \textbf{85}, 063406 (2012).

\bibitem[{\citenamefont{Herman}(1993)}]{Herman:CP:1984}
  \bibinfo{author}{\bibfnamefont{M.~F.}~\bibnamefont{Herman}} \bibnamefont{and}
  \bibinfo{author}{\bibfnamefont{E.}~\bibnamefont{Kluk}},
  \bibinfo{journal}{Chem. Phys.} \textbf{\bibinfo{volume}{91}},
  \bibinfo{pages}{27} (\bibinfo{year}{1984}).

\bibitem[{\citenamefont{Shalashilin}(2000)}]{Dmitry:JCP:2000}
  \bibinfo{author}{\bibfnamefont{D.~V.} \bibnamefont{Shalashilin}} \bibnamefont{and}
  \bibinfo{author}{\bibfnamefont{M.~S.}~\bibnamefont{Child}},
  \bibinfo{journal}{J. Chem Phys.} \textbf{\bibinfo{volume}{113}},
  \bibinfo{pages}{10028} (\bibinfo{year}{2000}).

\bibitem[{\citenamefont{Burnett et~al.}(1992)\citenamefont{Burnett, Reed,
  Cooper, and Knight}}]{burnett:PRA:1992}
\bibinfo{author}{\bibfnamefont{K.}~\bibnamefont{Burnett}},
  \bibinfo{author}{\bibfnamefont{V.~C.} \bibnamefont{Reed}},
  \bibinfo{author}{\bibfnamefont{J.}~\bibnamefont{Cooper}}, \bibnamefont{and}
  \bibinfo{author}{\bibfnamefont{P.~L.} \bibnamefont{Knight}},
  \bibinfo{journal}{Phys. Rev. A} \textbf{\bibinfo{volume}{45}},
  \bibinfo{pages}{3347} (\bibinfo{year}{1992}).

\bibitem{Krause:PRA:1992} J. L. Krause, K. J. Schafer, and K. C. Kulander, Phys. Rev. A \textbf{45}, 4998 (1992)

\bibitem[{\citenamefont{Bohm}(1952{\natexlab{a}})}]{bohm:PR:1952-1}
\bibinfo{author}{\bibfnamefont{D.}~\bibnamefont{Bohm}}, \bibinfo{journal}{Phys.
  Rev.} \textbf{\bibinfo{volume}{85}}, \bibinfo{pages}{166} (\bibinfo{year}{1952}); ibid. \bibinfo{pages}{180}
  (\bibinfo{year}{1952}{\natexlab{a}}).

%
\bibitem[{\citenamefont{Holland}(1993)}]{holland-bk}
\bibinfo{author}{\bibfnamefont{P.~R.} \bibnamefont{Holland}},
  \emph{\bibinfo{title}{The Quantum Theory of Motion}}
  (\bibinfo{publisher}{Cambridge University Press},
  \bibinfo{address}{Cambridge}, \bibinfo{year}{1993}).

\bibitem[{\citenamefont{Lai et~al.}(2009{\natexlab{a}})\citenamefont{Lai, Cai,
  and Zhan}}]{lai:EPJD:2009}
\bibinfo{author}{\bibfnamefont{X.~Y.} \bibnamefont{Lai}},
  \bibinfo{author}{\bibfnamefont{Q.~Y.} \bibnamefont{Cai}}, \bibnamefont{and}
  \bibinfo{author}{\bibfnamefont{M.~S.} \bibnamefont{Zhan}},
  \bibinfo{journal}{Eur. Phys. J. D} \textbf{\bibinfo{volume}{53}},
  \bibinfo{pages}{393} (\bibinfo{year}{2009}{\natexlab{a}});
  \bibinfo{journal}{New J. Phys.} \textbf{\bibinfo{volume}{11}},
  \bibinfo{pages}{113035} (\bibinfo{year}{2009}{\natexlab{b}}); \bibinfo{journal}{Chin. Phys. B} \textbf{\bibinfo{volume}{19}},
  \bibinfo{pages}{020302} (\bibinfo{year}{2010}).

\bibitem[{\citenamefont{Botheron and
  Pons}(2010{\natexlab{a}})}]{botheron:PRA-2:2010}
\bibinfo{author}{\bibfnamefont{P.}~\bibnamefont{Botheron}} \bibnamefont{and}
  \bibinfo{author}{\bibfnamefont{B.}~\bibnamefont{Pons}},
  \bibinfo{journal}{Phys. Rev. A} \textbf{\bibinfo{volume}{83}},
  \bibinfo{pages}{062704} (\bibinfo{year}{2011}{\natexlab{a}}).

\bibitem[{\citenamefont{Takemoto and Becker}(2011)}]{takemoto:JCP:2011}
\bibinfo{author}{\bibfnamefont{N.}~\bibnamefont{Takemoto}} \bibnamefont{and}
  \bibinfo{author}{\bibfnamefont{A.}~\bibnamefont{Becker}},
  \bibinfo{journal}{J. Chem. Phys.} \textbf{\bibinfo{volume}{134}},
  \bibinfo{pages}{074309} (\bibinfo{year}{2011}).

\bibitem{mompart}
 A. Pic\'on {\it et al.}, New J. Phys {\bf 12}, 083053 (2010).


\bibitem[{\citenamefont{Botheron and
  Pons}(2010{\natexlab{b}})}]{botheron:PRA-1:2010}
\bibinfo{author}{\bibfnamefont{P.}~\bibnamefont{Botheron}} \bibnamefont{and}
  \bibinfo{author}{\bibfnamefont{B.}~\bibnamefont{Pons}},
  \bibinfo{journal}{Phys. Rev. A} \textbf{\bibinfo{volume}{82}},
  \bibinfo{pages}{021404R} (\bibinfo{year}{2010}{\natexlab{b}}).

\bibitem{Song_2012} Y. Song, F. M. Guo, S. Y. Li, J. G. Chen, S. L. Zeng, Y. J. Yang, Phys. Rev. A \textbf{86}, 033424 (2012).

\bibitem{Wu}
 J. Wu, A. S. Sanz, B. B. Augstein, and C. Figueira de Morisson Faria
 (submitted for publication), pre-print arXiv:1301.1916

\bibitem{note-wyatt}
 One can also proceed the other way around,
 i.e., devising numerical methodology aimed at directly obtaining the
 Bohmian trajectories, from which relevant information about the
 quantum system is extracted, as done with the so-called {\it quantum
 trajectory methods} [e.g., see: R.E. Wyatt, {\it Quantum Dynamics
 with Trajectories} (Springer, New York, 2005)].

 \bibitem{timefrequency2}
 Ph. Antoine, B. Piraux, and A. Maquet, Phys. Rev. A \textbf{51},
 1750 (1995).

\bibitem{FDS1997}
 C. Figueira de Morisson Faria, M D\"orr, and W. Sandner,
 Phys. Rev. A \textbf{55}, 3961 (1997).

\bibitem{Belgium1998}
 A. de Bohan, Ph. Antoine, D. B. Milo\v{s}evi\'{c}, and B. Piraux,
 Phys. Rev. Lett. \textbf{81}, 1837 (1998).

\bibitem{timefrequency1}
 X.-M. Tong and S.-I. Chu, Phys. Rev. A \textbf{61}, 021802(R)(2000).
 \bibitem{Ruggenthaler_2008} M. Ruggenthaler, S. V. Popruzhenko, and D. Bauer, Phys. Rev. A \textbf{78}, 033413 (2008).
 \bibitem[{\citenamefont{Lein}(2010)}]{Lein:PRA:2010}
\bibinfo{author}{\bibfnamefont{C.~C.} \bibnamefont{Chiril${\breve{a}}$} \bibnamefont{{\it et al.}}},
  \bibinfo{journal}{Phys. Rev. A} \textbf{\bibinfo{volume}{81}},
  \bibinfo{pages}{033412} (\bibinfo{year}{2010}).
 \bibitem{Ciappina_1_2012} M. F. Ciappina, J. Biegert, R. Quidant, and M. Lewenstein, Phys. Rev. A \textbf{85}, 033828 (2012)
 \bibitem{Ciappina_2_2012} M. F. Ciappina, S. S. A\'{c}imovi\'{c}, T. Shaaran, J. Biegert, R. Quidant, and M. Lewenstein, Opt. Express \textbf{24}, 26261 (2012).


\bibitem[{\citenamefont{Knight}(1996)}]{Knight:PRA:1996}
\bibinfo{author}{\bibfnamefont{M.} \bibnamefont{Protopapas} \bibnamefont{{\it et al.}}},
  \bibinfo{journal}{Phys. Rev. A} \textbf{\bibinfo{volume}{53}},
  \bibinfo{pages}{R2933} (\bibinfo{year}{1996}).

\bibitem[{\citenamefont{Rost}(1999)}]{Rost:PRL:1999}
\bibinfo{author}{\bibfnamefont{G. van de} \bibnamefont{Sand}} \bibnamefont{and}
  \bibinfo{author}{\bibfnamefont{J.~M.} \bibnamefont{Rost}},
  \bibinfo{journal}{Phys. Rev. Lett.} \textbf{\bibinfo{volume}{83}},
  \bibinfo{pages}{524} (\bibinfo{year}{1999}).

 \bibitem[{\citenamefont{Carlos}(2012)}]{Carlos:PRA:2012}
\bibinfo{author}{\bibfnamefont{C.} \bibnamefont{Zagoya} \bibnamefont{{\it et al.}}},
  \bibinfo{journal}{Phys. Rev. A} \textbf{\bibinfo{volume}{85}},
  \bibinfo{pages}{041401(R)} (\bibinfo{year}{2012}).

\bibitem{Jie:Conference:2012} J. Wu {\it et al.},  ``Coupled-coherent State Approach for High-order Harmonic Generation", {\it High Intensity Lasers and High Field Phenomena conference, Berlin, Germany,
(2012)}.


\bibitem{Antoine_1996} P. Antoine, A. L'Huillier, and M. Lewenstein, Phys. Rev. Lett. \textbf{77}, 1234 (1996).

\bibitem{Hostetter_2010} J. A. Hostetter, J. L. Tate, K. J. Schafer, and M. B. Gaarde, Phys. Rev. A \textbf{82}, 023401 (2010)

\bibitem{Andreas} S. Dey and A. Fring, arXiv:1305.4619.


\bibitem{Perez_2010} J. A. P\'{e}rez-Hernandez, J. Ramos, L. Roso and L. Plaja, Laser Phys. \textbf{20}, 1044 (2010).

\bibitem[{\citenamefont{Gaarde}(2002)}]{Gaarde:PRA:2002}
\bibinfo{author}{\bibfnamefont{M.~B.} \bibnamefont{Gaarde} \bibnamefont{{\it et al.}}},
  \bibinfo{journal}{Phys. Rev. A} \textbf{\bibinfo{volume}{65}},
  \bibinfo{pages}{031406(R)} (\bibinfo{year}{2002}).

\end{thebibliography}

\end{document}